\documentclass[]{spie}  

\usepackage{amsmath,amsfonts,amssymb}
\usepackage{graphicx}
\usepackage[colorlinks=true, allcolors=blue]{hyperref}
\usepackage{comment}

\title{An innovative architecture for a wide band transient monitor on board the HERMES nano-satellite constellation}

\author[a,b]{F.~Fuschino}
\author[a,b]{R.~Campana}
\author[a,b]{C.~Labanti}
\author[c,d]{Y.~Evangelista}
\author[e]{F.~Fiore}
\author[f]{M.~Gandola}
\author[g]{M.~Grassi}
\author[f]{F.~Mele}
\author[c]{F.~Ambrosino}
\author[c]{F.~Ceraudo}
\author[h]{E.~Demenev}
\author[l]{M.~Fiorini}
\author[a]{G.~Morgante}
\author[c]{R.~Piazzolla}
\author[f]{G.~Bertuccio}
\author[g]{P.~Malcovati}
\author[h,m]{P.~Bellutti}
\author[h,m]{G.~Borghi}
\author[o,n]{G.~Dilillo}
\author[c,d]{M.~Feroci}
\author[h,m]{F.~Ficorella}
\author[i]{G.~La~Rosa}
\author[i]{P.~Nogara}
\author[o,n]{G.~Pauletta}
\author[h,m]{A.~Picciotto}
\author[m]{I.~Rashevskaya}
\author[n]{A.~Rachevski}
\author[i]{G.~Sottile}
\author[o,n]{A.~Vacchi}
\author[a]{E.~Virgilli}
\author[n]{G.~Zampa}
\author[n]{N.~Zampa}
\author[h,m]{N.~Zorzi}
\author[p]{T.~Chen}
\author[p]{N.~Gao}
\author[p]{J.~Cao}
\author[p]{Y.~Xu}
\author[p]{L.~Wang}

\affil[a]{INAF/OAS, Via Gobetti 101, I-40129, Bologna, Italy}
\affil[b]{INFN--Sezione di Bologna, Viale Berti Pichat 6/2, I-40127, Bologna. Italy}
\affil[c]{INAF/IAPS Roma, Via Fosso del Cavaliere 100, I-00133, Roma, Italy}
\affil[d]{INFN-Tor Vergata, Via della Ricerca Scientifica, 1 I-00133,  Roma - Italy}
\affil[e]{INAF/OATS, Via G.B. Tiepolo 11, I-34143, Trieste, Italy} 
\affil[f]{Department of Electronics, Information and Bioengineering (DEIB) of Politecnico di Milano, Como Campus, Via Anzani 42, 22100 Como, Italy}
\affil[g]{Department of Electrical, Computer, and Biomedical Engineering, University of Pavia, Via Ferrata 5, 27100, Pavia, Italy}
\affil[h]{Fondazione Bruno Kessler – FBK, Via Sommarive 18, I-38123 Trento, Italy}
\affil[i]{INAF, Istituto di Astrofisica Spaziale e Fisica cosmica di Palermo, via U. La Malfa 153, I-90146 Palermo, Italy}
\affil[l]{INAF-IASF Milano, Via Alfonso Corti 12, I-20133 Milano, Italy} 
\affil[m]{TIFPA-INFN, Via Sommarive 14, I-38123 Trento, Italy}
\affil[n]{INFN Italian National Institute for Nuclear Physics c/o Area di Ricerca Padriciano 99, I-34127 Trieste, Italy}
\affil[o]{Department of Mathematics, Computer Science and Physics University of Udine, Via delle Scienze 206, I-33100 Udine, Italy}
\affil[p]{Institute of High Energy Physics, Chinese Academy of Sciences, Beijing, China}

\authorinfo{Send correspondence to:\\
Fabio Fuschino, E-mail: \url{fabio.fuschino@inaf.it}\\
}

\begin{document} 
\maketitle

\begin{abstract}
The HERMES-TP/SP mission, based on a nanosatellite constellation, has very stringent constraints of sensitivity and compactness, and requires an innovative wide energy range instrument. The instrument technology is based on the “siswich” concept, in which custom-designed, low-noise Silicon Drift Detectors are used to simultaneously detect soft X-rays and to readout the optical light produced by the interaction of higher energy photons in GAGG:Ce scintillators. To preserve the inherent excellent spectroscopic performances of SDDs,  advanced readout electronics is necessary. In this paper, the HERMES detector architecture concept will be described in detail, as well as the specifically developed front-end ASICs (LYRA-FE and LYRA-BE) and integration solutions. The experimental performance of the integrated system composed by scintillator+SDD+LYRA ASIC will be discussed, demonstrating that the requirements of a wide energy range sensitivity, from 2 keV up to 2 MeV, are met in a compact instrument.
\end{abstract}


\section{INTRODUCTION}
\label{intruduction}

Gamma-Ray Bursts (GRBs) are one of the most intriguing
and challenging phenomena for modern science, because of their huge luminosities, up to more than $10^{52}$~erg/s, their redshift distribution extending from $z\sim~0.01$ up to $z > 9$. Their study is of very high interest for several fields of astrophysics, such as the physics of matter in extreme conditions, cosmology, fundamental physics and the mechanisms of gravitational 
wave signal production.

The \emph{High Energy Modular Ensemble of Satellites - Technologic and Scientific Pathfinder} (HERMES-TP/SP) \cite{fuschino19} is a new mission concept hosting compact but innovative X-ray detectors for the monitoring of cosmic high energy transients, such as Gamma Ray Bursts and the electromagnetic counterparts of Gravitational Wave events, based on a constellation of six 3U nanosatellites. The main topic of HERMES is to demonstrate that accurate location of high energy cosmic transients can be obtained using CubeSats. That means miniaturized hardware, one order of magnitude smaller costs than conventional space observatories and few years of development time. The four pillars of the HERMES project are: 1) providing a first mini-constellation for GRB localization with a total of six units, for GRB triangulation using miniaturized and distributed instrumentation, joining also to the multimessenger revolution; 2) develop miniaturized payload technology for breakthrough science; 3) contribute to the so-called Space 4.0 goals demonstrating that components off-the shelf (COTS) can be reliable for challenging missions; 4) push and prepare for final and more solid larger constellation.

The HERMES-TP project is funded by the Italian Ministry for education, university and research and the Italian Space Agency. The HERMES-SP project is funded by the European Union’s Horizon 2020 Research and Innovation Programme under Grant Agreement No. 821896. The constellation should be tested in orbit in 2022. HERMES-TP/SP is intrinsically a modular experiment that can be naturally expanded to provide a global, sensitive all sky monitor for high energy transients.

\section{HERMES Mission Concept}
\label{general_overview}
The HERMES approach differs from the conventional idea to build increasingly larger and expensive instruments. HERMES proposes to realize distributed, modular and innovative instruments composed by tens/hundreds of simple units, cheaper and with a limited development time.
The current CubeSat technologies demonstrated that off-the-shelf components for space use can offer solid readiness at a limited cost.
The physical dimension of a single HERMES detector need to be compliant with the nanosatellite mechanical structure; the 1U  CubeSat standard is 10$\times$10$\times$10 cm$^3$. Therefore, the single HERMES detector is of course under-performing, having a lower effective area when compared with transient monitors currently operating. The lower costs and the innovative concept of distributed instrument demonstrate that to build an new mission with unprecedented sensitivity is actually feasible. 
The HERMES detector is designed to provide a sensitive area $>$50 cm$^2$, suggesting that with several tens/hundreds of such units a total sensitive area of the order of magnitude of $\sim$1 m$^2$ can be reached.
By measuring the time delay between different satellites, using cross correlation of the lightcurves detected by different satellites \cite{sanna20}, the localisation capability of the whole constellation is directly proportional to the number of components and inversely proportional to the average baseline between them.
As a rough example, with a reasonable average baseline of $\sim$7000 km, represented by the Earth radius, and a reasonable number for low-Earth satellites in suitable orbits (e.g. $\sim$100 nanosats) simultaneously detecting a transient, a source localisation accuracy of the order of magnitude of $\sim$10 arcsec$^2$ can be
reached, for transients with milliseconds time scale variability.

A more extensive description of HERMES mission concept is given in other proceedings in this volume \cite{fiore20, colagrossi20, sanna20}.

\section{Payload description}
\label{PL}
\begin{figure} [ht]
\centering
\includegraphics[height=7.cm]{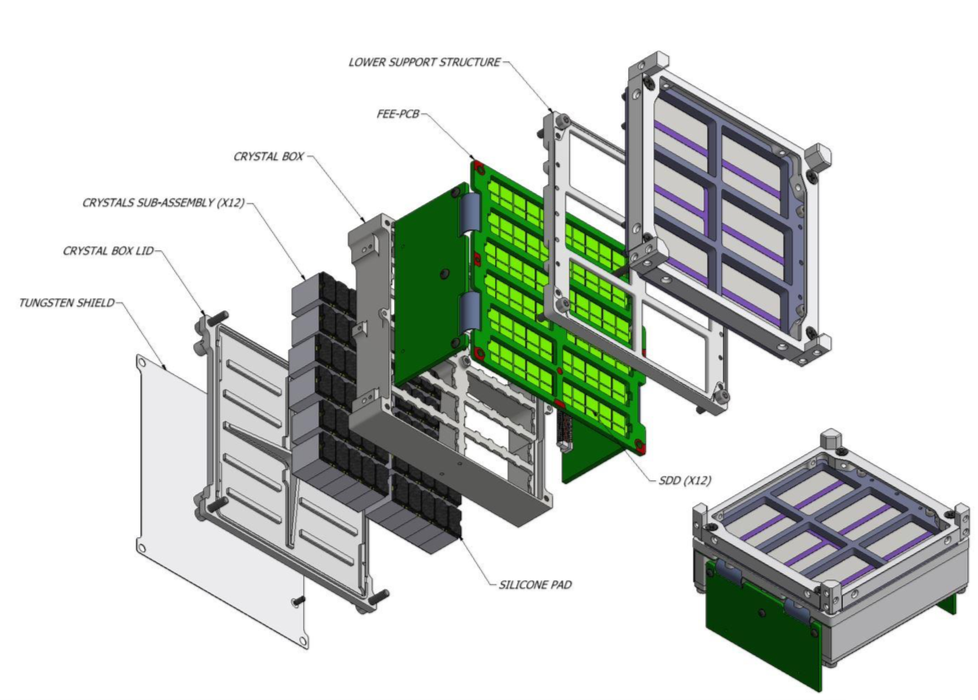}
\caption[example] 
{ \label{HERMESpl}
Exploded view of the payload unit (10$\times$10$\times$10 cm$^3$) on board the HERMES nanosatellite. From the top are shown the mechanical support composed by top and bottom parts, with optical filter (violet) in middle, the FEE board (dark green) allocating SDD matrices (light green), LYRA-FE chips on the top and LYRA-BE chips folded on the side (not shown for clarity), the GAGG:Ce crystal pixels and their housing. A mechanical rib on top is also visible, necessary to fix the whole payload to the satellite structure.}
\end{figure}

A sketch of the HERMES payload, allocated 
in 1U-CubeSat (10$\times$10$\times$10 cm$^3$), is shown in Figure~\ref{HERMESpl}. 
A mechanical support is placed on the instrument topside. The support is composed by two parts to hold an optical/thermal filter. 
The detector core is located under the mechanical support: this is a scintillator-based detector in which Silicon Drift Detectors (SDD\cite{gatti84}) are used to both detect soft X-rays, by direct absorption in silicon, and to simultaneously readout the scintillation light. 
The payload unit is expected to allocate a detector with $>$50 cm$^2$ sensitive area in the 
energy range from 2 keV up to 2 MeV, with a total power consumption $<$5 W and total weight of $\sim$1.5 kg.
The payload unit is completed by further electronic boards implementing back-end electronics, power supply and data handling functionalities. They are allocated on the bottom of the payload unit, below the detector unit.
A more extensive description of HERMES payload is given in other proceedings in this volume \cite{evangelista20}.

\section{Detector core architecture}
\label{DetectorARC}
The HERMES detector is based on the so-called “siswich” concept \cite{marisaldi04,marisaldi08}, exploiting the dual intrinsic sensitivity of silicon to both soft X-rays and scintillation light. Thus, aiming at designing a compact instrument with a very wide sensitivity band, the silicon detectors are optically coupled with inorganic scintillators. The detector is therefore composed by an array of scintillator pixels, optically insulated, read out by SDDs.

\begin{figure} [ht]
\centering
\includegraphics[height=10cm]{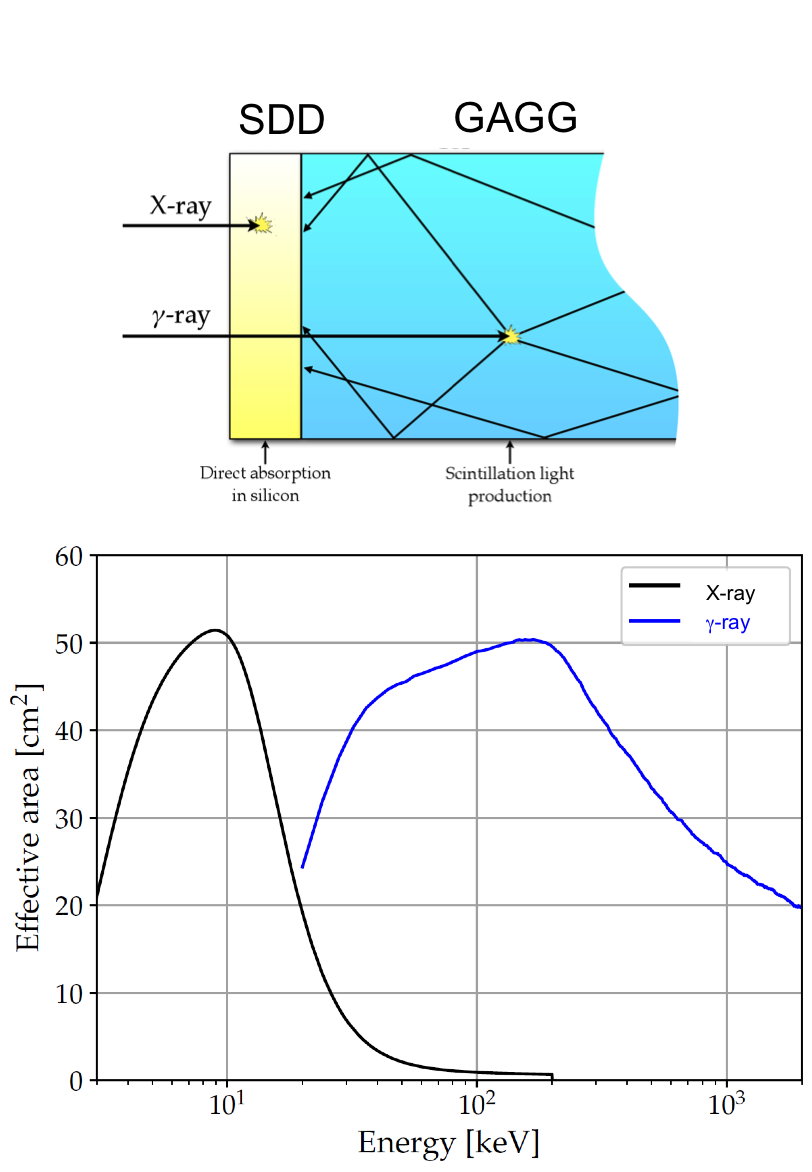}
\caption[example] 
{ \label{eff}
\emph{Top:} the working principle of a \emph{siswich} detection system. The SDD, directly facing the sky, is coupled with a GAGG:Ce scintillator crystal. The low energy radiation (below $\sim$30 keV) interacts in the SDD. Photons with higher energy cross the SDD and interact in the scintillator. The light output is collected by SDD.  \emph{Bottom:} efficiency vs. photon energy for the direct detection in the SDD (black curve) and through light conversion in the GAGG:Ce (Blue curve). }
\end{figure}

In this concept the SDDs play also the role of an independent X-ray solid state detector. When the SDD faces the sky, low energy X-rays are
directly absorbed by the SDD, while higher energy X-rays and
$\gamma$-rays, pass through the silicon sensor without interactions and are absorbed in the scintillator crystal. The optical scintillation photons are than collected by the same silicon sensor. Only very low noise readout devices coupled with state-of-the-art front-end electronics allow to reach a low energy scintillator threshold down to 20--30 keV. This is a key issue of this architecture, considering that around these energies the increasing sensitivity of the scintillator is able to compensate the lack of efficiency of thin silicon wafers (450 $\mu$m), so a quite flat overall sensitivity in a wide energy band for the whole integrated system is reached. This effect is clearly visible in the Figure~$\ref{eff}$ where both silicon and scintillator detection efficiency vs. photon energy are shown.

The inorganic scintillator selected for HERMES is the Cerium-doped Gadolinium-Aluminum-Gallium Garnet (GAGG:Ce) \cite{kamada12}. This is a very promising material with high light output ($\sim$50\,000 ph/MeV), a fast radiation decay time of $\sim$90 ns, no hygroscopicity, high density (6.63 g/cm$^3$) with an effective mean atomic number of 54.4, no internal radioactive background, peak light emission at 520 nm. These specifications make this material very suitable for the HERMES detector. Since GAGG:Ce is a relatively recent crystal, it has not been yet extensively investigated with respect to radiation resistance and performance after irradiation, although the published results are very encouraging \cite{dilillo20}.
Further radiation tests were carried out by our collaboration, to better understand the scintillator behaviour in specific in-orbit environments to evaluate their impact on the scientific performance. A more in-depth description of in-orbit conditions and irradiation campaign results are presented in other proceedings in this volume \cite{dilillo20,ripa20,campana20}.
These tests confirmed the very good performance of GAGG:Ce compared to other scintillator materials largely used in the recent years in space-borne experiments for $\gamma$-ray astronomy (e.g. BGO or CsI).

\section{The HERMES Silicon Drift Detectors}
\begin{figure} [ht]
\centering
\begin{tabular}{cc}
\includegraphics[height=4.cm]{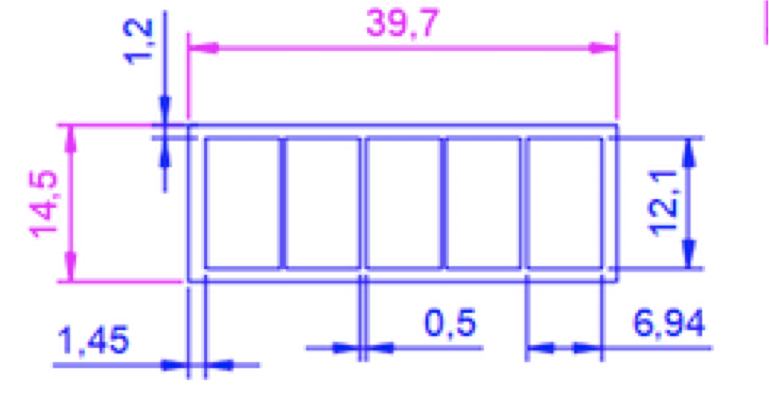}
\includegraphics[height=4.cm]{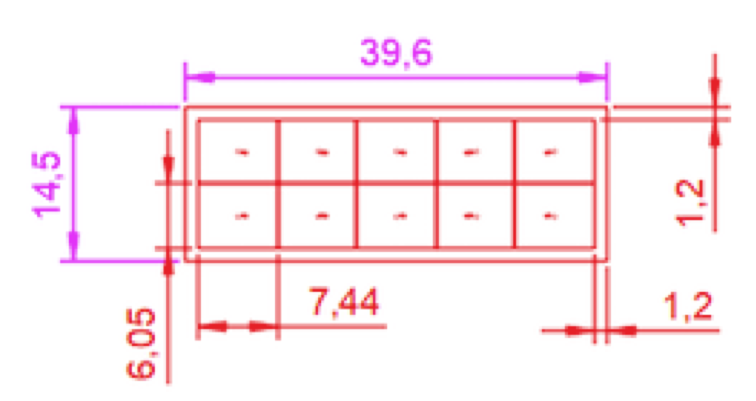}
\end{tabular}
\caption[example] 
{ \label{sdd}
SDD design. Left: $p$-side, right: $n$-side}
\end{figure}

The SDD development builds on the state-of-the-art results achieved within the framework of the Italian ReDSoX collaboration\footnote{\url{http://redsox.iasfbo.inaf.it}}, with the combined design and manufacturing technology coming by a strong synergy between INFN-Trieste and Fondazione Bruno Kessler (FBK, Trento), in which both INFN and FBK co-fund the production of ReDSoX Silicon sensors. Thanks to several project funded by ASI, INAF and INFN a large amount of work was done in past years to consolidate the design of SDDs, as well as the production process, for different applications, for both on-ground and space-based experiments \cite{zampa11,campana11,Rachevski14,campana16,fuschino16,evangelista18,cirrincione19}.
The current state of the consolidated design/process concerns, for example, pitch and distance of the drift cathodes, low power consumption of sensors, extremely low dark current, high detection efficiency (both for ionizing radiation and optical photons), enhanced efficiency on the sensor edges, optimized charge collection efficiency.
The SDDs are the base elements around which the overall P/L is designed. The HERMES SDD design is shown in Figure~\ref{sdd}. Nominal thickness of the SDD arrays is 450 $\mu$m, with a certified standard deviation of 2.88 $\mu$m (by manufacturer).
The SDD is organized in a 2$\times$5 array. Guard rings surrounds the SDD matrix, allowing for a correct electric field termination. The guard ring area is passive, thus the overall geometric area of the SDD device is 39.6$\times$14.5 mm$^2$, while the sensitive area of each SDD cell is 6.05$\times$7.44 mm$^2$.
A single crystal ($\sim$12.1$\times$6.94 mm$^2$ and 15.0~mm thick) is coupled with two SDD channels on the $p$-side. On this side to avoid optical cross-talk coming from different crystals, a metal implant strip 0.5~mm wide is present between adjoining couple of cells (Figure~\ref{sdd} right panel).
With this configuration a signal from a single isolated SDD channel will be considered to be originated by the direct absorption of an X-ray in silicon, while a trigger (with comparable amplitude) from the two SDD channels coupled to the same crystal will be considered as a $\gamma$-ray event given the expected uniform illumination on both SDD cells by the scintillation light. 

\section{The HERMES readout ASIC: LYRA}

The potentially high energy resolution of SDDs can be reached only with highly custom and advanced front-end electronics requiring a preamplifier in close proximity to the anode \cite{bertuccio15}.
Being the HERMES detector constituted by 120 channels distributed over a total geometrical 
area $<$100~cm$^2$, a special architecture for the readout electronics is so required.
A low-noise, low-power Application Specific Integrated Circuit (ASIC) named LYRA has been conceived and designed for this task. The heritage for the LYRA ASICs comes from the VEGA project \cite{Aha14,campana14},
that was developed by Politecnico of Milano and University of Pavia for the readout of SDDs within the ReDSoX Collaboration during the LOFT Phase-A study (ESA M3 Cosmic Vision program), although a specific and renewed design is necessary to comply with the different SDD characteristics and the unique
system architecture in HERMES. 
Each LYRA chipset is divided into 30 Front-End ASIC (LYRA-FE) and 1 Back-End multichannel ASIC (LYRA-BE) for a total of 120 LYRA-FE and 4 LYRA-BE circuits.
The LYRA-FE chips include the charge sensitive amplifier (CSA), first shaping stage and a current-mode line driver. Each LYRA-BE channel is used to read and process the output of a single LYRA-FE, completing the spectroscopic chain with mixed-signal architecture \cite{gandola2019lyra}. 

\begin{figure} [ht]
\centering
\includegraphics[height=11.cm]{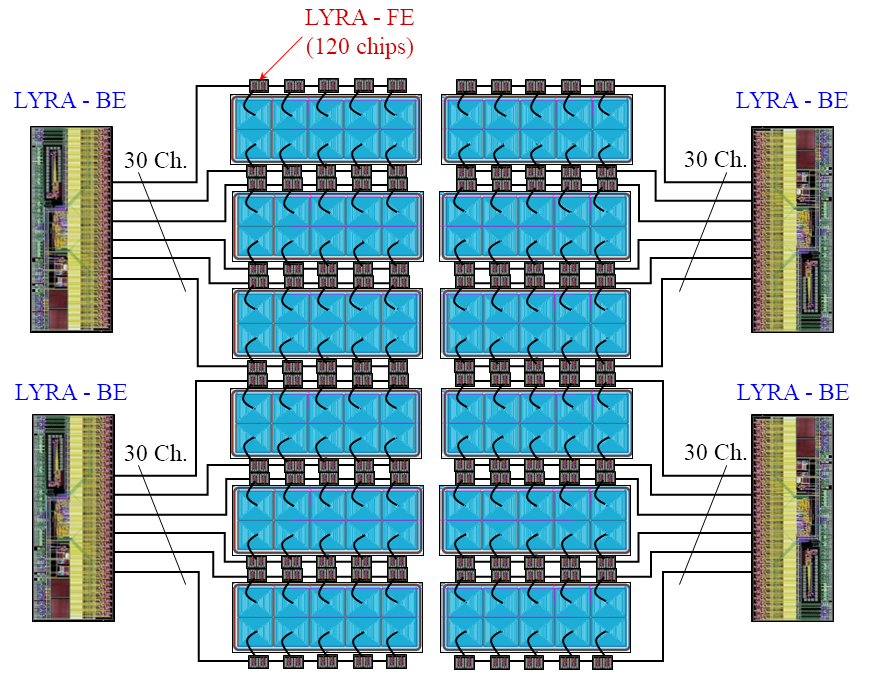}
\caption[example] 
{ \label{arcLYRA}
Architecture of the LYRA chipset with 120 LYRA-FE connected to 120 anodes of the SDD and 4 LYRA-BE.}
\end{figure} 

\subsection{LYRA-FE}

The LYRA-FE die is 0.9$\times$0.6 mm$^2$ and includes a complete back-up channel for offline replacement. Due to its compact dimensions the LYRA-FEs can be placed very close to the SDD anodes, in order to minimize the stray capacitance of the detector-preamplifier connection, improving the overall electronic noise performance. A total of 120 LYRA-FE chips are used to readout all the SDD channels in the HERMES detector. The LYRA-FEs transmit their analog output signals to the LYRA-BE ASIC in current-mode in order to inhibit the destructive inter-channel crosstalk that the classical voltage-mode signal transmission would cause. The LYRA-FE chip is shown in Figure~\ref{LYRA}. 

\subsection{LYRA-BE}

\begin{figure} [ht]
\centering
\begin{tabular}{cc}
\includegraphics[height=5.cm]{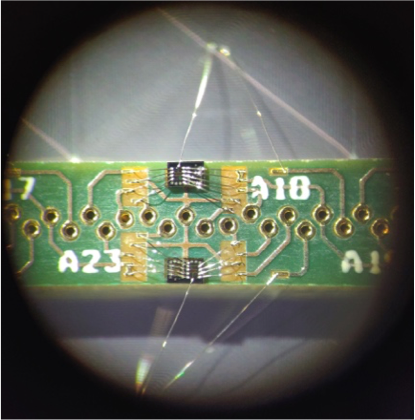}
\includegraphics[height=5.cm]{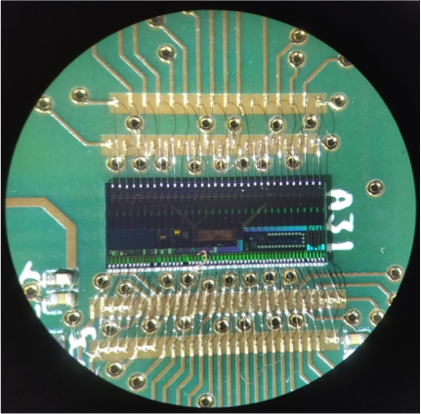}
\end{tabular}
\caption[example] 
{ \label{LYRA}
Microscope view of the LYRA-FE (left) and LYRA-BE (right) ASICs integrated on the FEE breadboard.}
\end{figure}

In the HERMES configuration, the LYRA-BE chips ($\sim$6.5$\times$2.5 mm$^2$ die), shown in Figure~\ref{LYRA}, can be placed out of the detection plane, where the SDD matrices and LYRA-FE chips are accommodated on a central rigid part. Two additional wings, where LYRA-BE are accommodated, are connected to the central part by means of embedded flex cables, which allow also avoiding the additional space required by connectors, offering the possibility to “fold” the wings at right angle with respect to the detection plane, on the external side of the payload unit, maximizing the detector effective area.
A total of 4 LYRA-BE chips are used for the whole payload. Each LYRA-BE has 30 operating channels in HERMES configuration, i.e. connected to LYRA-FE chips, and the individual LYRA-BEs are operated in parallel increasing also the redundancy of the system.
In detail a LYRA BE ASIC embeds 32 channels due to the presence of 2 back-up channels. Each channel is made up of a current-mode signal receiver, a second shaping stage, a discriminator, a peak stretcher circuit and a control logic. In addition the LYRA BE chip includes a configuration register and an output multiplexer  followed by an analog buffer to drive the input load on-board FPGA. Through the multiplexer all the analog outputs of each SDD (in shaper and in stretcher mode) can be read-out as well as the event detection triggers. The threshold for the discriminator can be programmed in two steps through the configuration serial register: a coarse value for all channel and a fine one for each pixel. For debug purpose all the pixels can be disabled by suitable configuration bits if required.

\section{Breadboard for cross-talk and preliminary performance measurements}

\begin{figure} [!ht]
\centering
\begin{tabular}{ccc}
\includegraphics[height=4.7 cm]{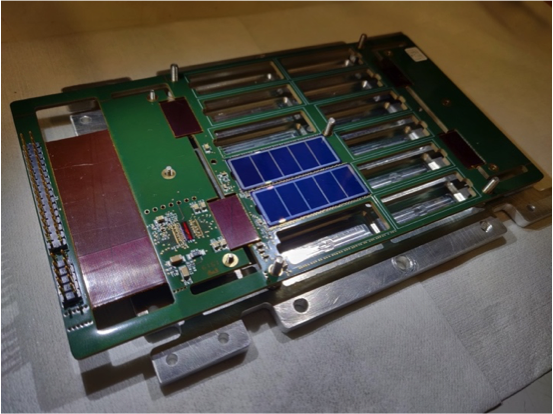}
\includegraphics[height=4.7 cm]{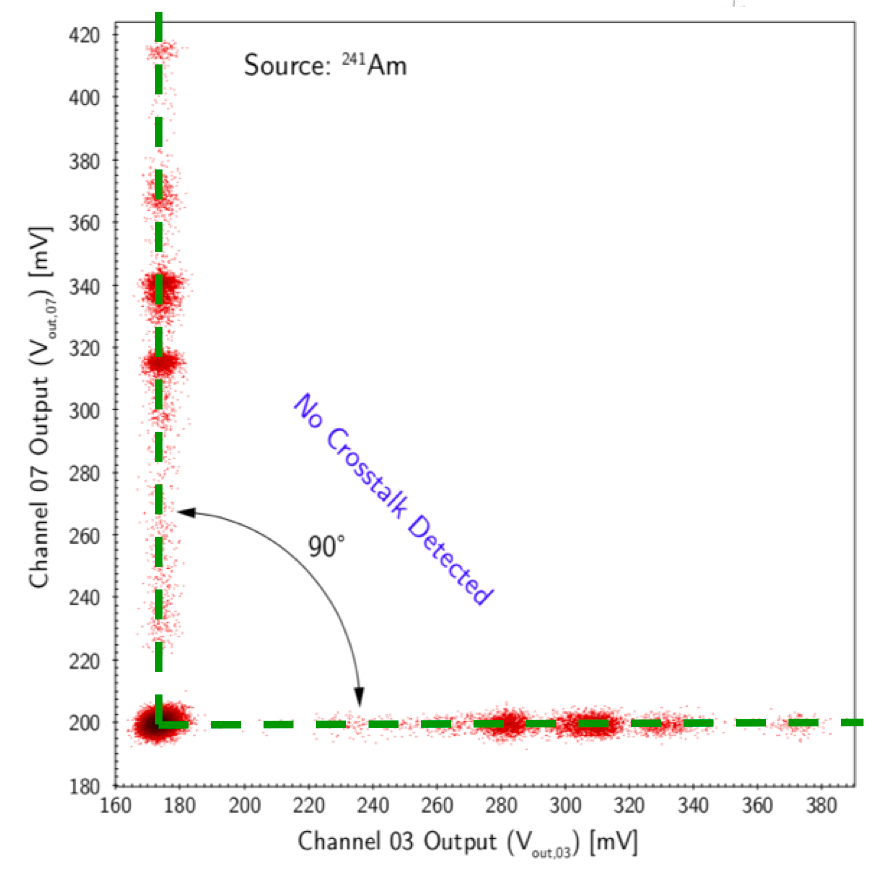}
\includegraphics[height=4.7 cm]{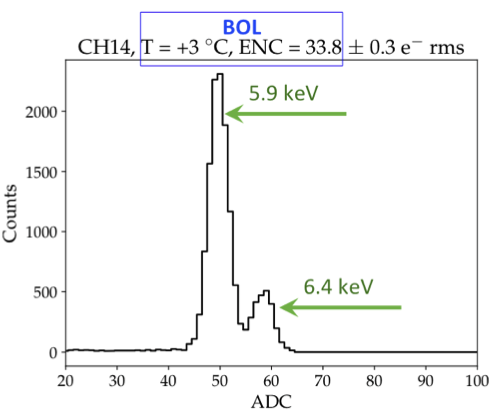}
\end{tabular}
\caption[example] 
{ \label{CrossTalk}
HERMES FEE breadboard (left), measured channel cross-talk (center) and energy spectrum acquired with a $^{55}$Fe radioisotope (right).
}
\end{figure} 

A FEE breadboard has been designed and manufactured to consolidate the PCB design verifying the system requirements in terms of performances. The rigid-flex technology used to cope with the strong space constraints imposed by the S/C, as well as the architecture of the readout electronics with two different ASIC exploiting current-based interfaces, represent the main innovative points of the final FEE PCB. An electrical characterization was therefore carried out to verify the absence of channel cross-talk that could degrade the energy resolution and the event discrimination performance.

Thanks to the modular design of the HERMES detector, the PCB layout of the test breadboard has been limited to the routing of a single quadrant (40 LYRA-FE dies) integrating 2 SDD detectors for a total of 20 SDD cells (Figure \ref{CrossTalk}, left). The channel cross-talk that could be introduced by the reduced distance of PCB traces carrying the LYRA-FE output analog signals has been characterized in INAF/OAS laboratories, in order to verify the PCB layout and to confirm current-based interface between the LYRA-FE analog outputs and the LYRA-BE analog inputs.
Central panel of Figure~\ref{CrossTalk} shows the scatter plot of the non-correlated (\textsuperscript{241}Am radioactive source) on two channels with adjacent copper traces in the PCB. The absence of systematic correlations of the two signals (shown as a perfect orthogonality between the signals generated by the two channels) demonstrates the negligibility of the cross-talk contribution to the system performance. Right panel of Figure~\ref{CrossTalk} shows instead a $^{55}$Fe spectrum acquired with the integrated breadboard in the INAF/OAS climatic chamber at a detector temperature of +3 $^\circ$C. The energy resolution at the 5.9 keV peak is 311 eV FWHM.
A detailed description of simulation and experimental results can be found in \cite{gandola2019lyra, grassi2020x}.

\section{The Demonstration Model: the first integrated complete HERMES detector}

\begin{figure} [ht]
\centering
\begin{tabular}{ccc}
\includegraphics[height=4.3cm]{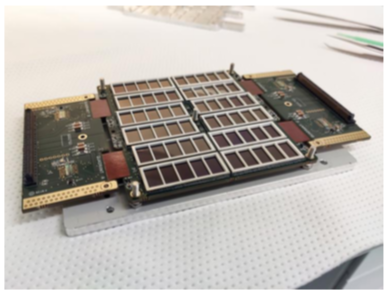}
\includegraphics[height=4.3cm]{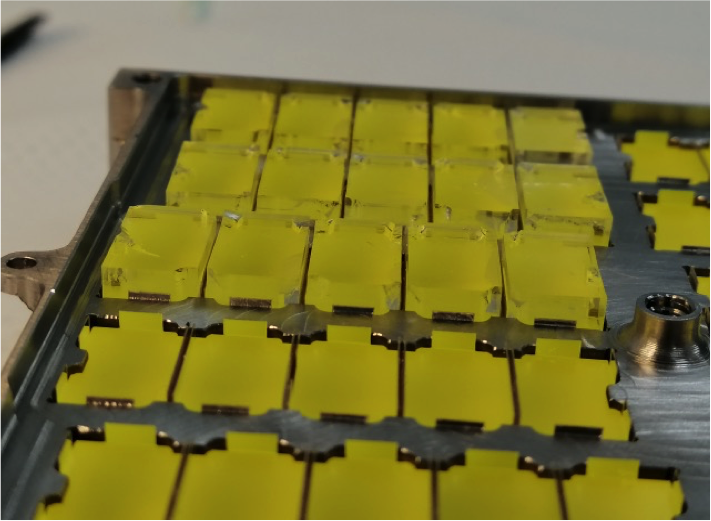}
\includegraphics[height=4.3cm]{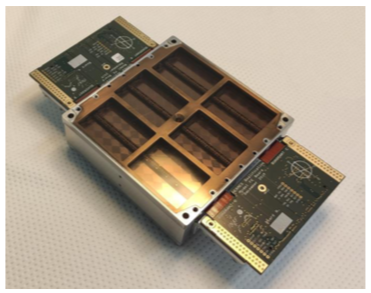}
\end{tabular}
\caption[example] 
{\label{detector} 
HERMES Detector Assembly workflow. The whole integrated FEE PCB (left) includes 120 LYRA-FEs, 12 SDD matrix and 4 LYRA-BEs; the crystal box (middle) contains 60 wrapped GAGG:Ce crystals, optical pads are also visible; whole integrated Detector Assembly (right) include a dummy optical filter, which is also visible. }
\end{figure}

Due to the required readiness level and the reduced development time, in the development plan of HERMES project the conventional model philosophy was customised, introducing what we called \emph{Payload Demonstration Model} (DM). This represent the first completely integrated non-flight HERMES payload, to be dedicated to validate the integration procedure, and to functional and preliminary performance tests, including verification/qualification tests. The DM is described more in detail in other proceedings in this volume \cite{evangelista20}.
The HERMES payload integration workflow is summarised in the Figure~\ref{detector}, where the whole integrated FEE PCB is shown on the left, the crystal box with the individual wrapped crystals in the middle and whole integrated Detector Assembly on the right.

Although the P/L DM is a non-flight element, it has been integrated with a representative set of flight-like components, following the finalized AIT/V procedures. 
The whole integrated Detector Assembly is composed by:
\begin{itemize}
    \item Mechanical assembly;
    \item Optical filter;
    \item FEE board equipped with:
    \begin{itemize}
        \item 4 LYRA-BE ASICs (one per DA quadrant);
        \item 120 LYRA-FE ASICs (30 per quadrant);
        \item 12 SDD arrays (4 in-spec + 8 dummy, mechanically representative);
    \end{itemize}
    \item 60 GAGG:Ce crystals;
    \item Fully representative optical coupling and crystal preload pads;
    \item Tungsten shields;
\end{itemize}

Because of the mechanical fragility of the quite thin SDDs (450 $\mu$m) combined with soft optical coupling pads between crystals and SDD matrix and the PCB support for Silicon sensor only on the surrounding perimeter, an additional frame (ABS+Kovar)
is glued on the $p$-side of SDD for mechanical strenghtening. The frame geometry is also optimised to minimize potential optical cross-talk between contiguous crystals. This frame (white) is clearly visible in the Figure~\ref{detector}, left.
Here the reinforced SDD are glued on the bottom side of FEE PCB (in the Figure~\ref{detector} left, the FEE PCB is up side down), facing the frame side towards scintillators. For the optical coupling a soft silicone pad (one per crystal) is used. To facilitate the coupling of 60 crystals and 12 SDD matrices, wrapped scintillator crystals are arranged in the so called crystal box (Figure~\ref{detector}, center).
This box is optimised to precisely couple crystals and controlling also the compression against Silicon sensors.
Completing the integration of Detector Assembly is the mechanical structure on the top of FEE PCB, a two-component frame designed to fix FEE PCB in a "sandwich" configuration towards the crystal box. The top mechanical structure is able also to accommodate an optical filter. A dummy optical filter is visible in the Figure~\ref{detector} right.

\section{Detector Assembly: preliminary performance evaluation}

\begin{figure} [ht]
\centering
\begin{tabular}{cc}
\includegraphics[height=5cm]{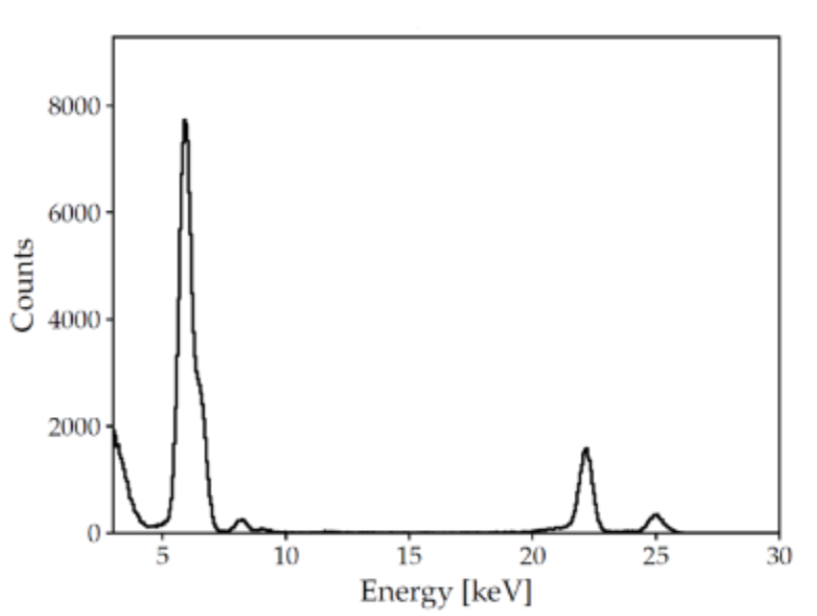}
\includegraphics[height=5cm]{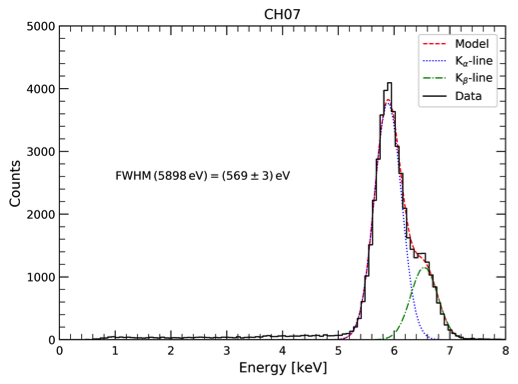}
\end{tabular}
\caption[example] 
{\label{spectra} 
Left: $^{55}$Fe and $^{109}$Cd spectrum acquired on one detector channel at a temperature of (26$\pm$1) $^\circ$C. Right: $^{55}$Fe spectrum acquired at $T=26\pm1$ $^\circ$C, where the Mn~K$_\alpha$ and Mn~K$_\beta$ best-fit was used to estimate the overall noise.
}
\end{figure}

In order to assess the X-rays and $\gamma$-rays detection capabilities and preliminary performance, the integrated Detector Assembly (DA) has been characterized with radioactive sources at the IAPS Rome facility. The DA has been powered by means of unfiltered laboratory power supplies connected, through an adapter board, directly to the LYRA-BE and LYRA-FE ASIC and to the SDD bias network. The LYRA-BE ASICs command and communication tasks have been managed through a custom Test Equipment including a  commercial multi-channel fast ADC (CAEN DT5740) for data acquisition. 
The Detector Assembly has been characterized with X-ray sources at the IAPS laboratories. A series of energy spectra have been acquired at +26 $^\circ$C and +5 $^\circ$C by illuminating the detector with $^{55}$Fe (Mn~K$_\alpha$ at 5.9 keV, Mn~K$_\beta$ at 6.49 keV) and $^{109}$Cd (Ag~K$_\alpha$ at 22.1 keV, Ag~K$_\beta$ at 24.9 keV). In Figure~\ref{spectra} we show an example of the collected X-ray spectra: on the left a raw spectrum, i.e. without any selection between X-rays and $\gamma$-rays, acquired stimulating  the detector simultaneously with $^{55}$Fe and $^{109}$Cd sources at 26 $^\circ$C, while on the right panel a $^{55}$Fe spectrum is shown.

\begin{figure} [ht]
\centering
\includegraphics[height=5cm]{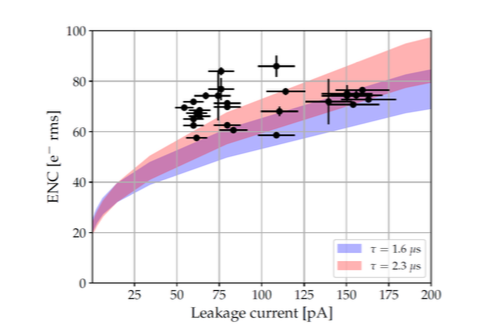}
\caption[example] 
{\label{perf} 
Measured $I_\mathrm{leak}$-ENC values (black circles) at $T=26\pm1$ $^\circ$C superimposed to the simulated system performance. See text for more details.
}
\end{figure}

The characterization of the integrated DA included the comparison between the leakage current measurement for all SDD cells and the final noise performances estimated by $^{55}$Fe spectra. Using the $I_\mathrm{leak}$ values, measured for each SDD cell, and the noise values estimated by X-rays spectra, we verified the $I_\mathrm{leak}$-ENC relationship obtained for the SDD+LYRA system by means of simulations.
Figure~\ref{perf} shows the measured $I_\mathrm{leak}$-ENC values for each channel (black circles), superimposed to the simulated system performance for the two selectable peaking times (red and blue areas, $\tau_\mathrm{peak}$=1.6 $\mu$s and 2.3 $\mu$s respectively). Although a set-up induced noise excess ($\sim$10 e$^-$ r.m.s) is evident for low detector leakage currents, for $I_\mathrm{leak} > 90$~pA the experimental data match the expected system performance. Noise performance with a flight-representative set-up will be assessed after the electrical integration of the power-supply unit (PSU) PCB with the DA.
Using the current temporary set-up, an average value $<$0.8 keV for X-rays low energy threshold was calculated on a total of 40 characterized SDD channels. 

\begin{figure} [ht]
\centering
\includegraphics[height=7cm]{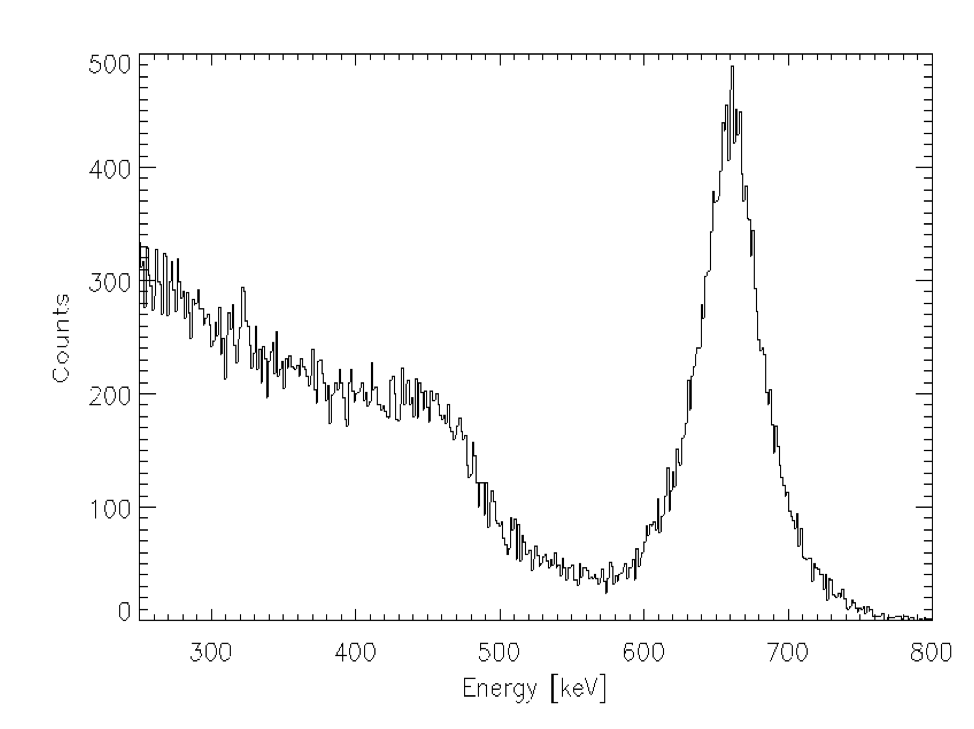}
\caption[example] 
{\label{cs} 
Typical $^{137}$Cs spectrum acquired on a single detector channel at a temperature of $T=26\pm1$ $^\circ$C, used to estimate $\gamma$-ray detection capabilities and performance.
}
\end{figure}

A $^{137}$Cs radioactive source ($\gamma$-ray at 662 keV) has been used to characterize the $\gamma$-ray detection capabilities and the scintillation light output response. 
A series of energy spectra have been acquired at +26 $^\circ$C and +5 $^\circ$C in order also to evaluate the increase in the light output with decreasing temperature. An average value of $\sim$12.2 e$^-$/keV was measured at 26$^\circ$C, when a signal from a single SDD cell is acquired, while an average increment of ~11.8\% in the light output is observed at a temperature of +5 $^\circ$C, demonstrating a full agreement with both design expectation and requirements as well as literature results. A typical $^{137}$Cs spectrum acquired during the HERMES Detector Assembly characterization is shown in Figure~\ref{cs}.

\section{Conclusions}
The experimental activities described in this paper, concerning the integration of the HERMES Detector Assembly, were carried out to successfully conclude, in these last weeks (end of November 2020), the Critical Design Review milestone. In this scenario, the experimental measurements described in the previous sections were used to demonstrate the satisfaction of requirements for both beginning-of-life, corresponding roughly to the measurements at +5~$^\circ$C, and end-of-life conditions, corresponding to the +26~$^\circ$C measurements, given the worsening of performance induced on the silicon sensor due to radiation damage during in-orbit life.

As an outcome of these activities, the HERMES wide energy range compact detector, sensitive from 2 keV up to 2 MeV, is now ready for all the necessary forthcoming steps of the System Assembly, Integration and Test of its Flight versions, in the Phase D of the project.

\acknowledgments 
This project has received funding from the European Union Horizon 2020 Research and Innovation Framework Programme under grant agreement HERMES-Scientific Pathfinder n. 821896 and from ASI-INAF Accordo Attuativo HERMES Technologic Pathfinder n. 2018-10-HH.0.

\bibliography{report} 

\begin{thebibliography}{10}

\bibitem{fuschino19}
{Fuschino}, F., {Campana}, R., {Labanti}, C., {Evangelista}, Y., {Feroci}, M.,
  {Burderi}, L., {Fiore}, F., {Ambrosino}, F., {Baldazzi}, G., {Bellutti}, P.,
  {Bertacin}, R., {Bertuccio}, G., {Borghi}, G., {Cirrincione}, D., {Cauz}, D.,
  {Ficorella}, F., {Fiorini}, M., {Gandola}, M., {Grassi}, M., {Guzman}, A.,
  {Rosa}, G.~L., {Lavagna}, M., {Lunghi}, P., {Malcovati}, P., {Morgante}, G.,
  {Negri}, B., {Pauletta}, G., {Piazzolla}, R., {Picciotto}, A., {Pirrotta},
  S., {Pliego-Caballero}, S., {Puccetti}, S., {Rachevski}, A., {Rashevskaya},
  I., {Rignanese}, L., {Salatti}, M., {Santangelo}, A., {Silvestrini}, S.,
  {Sottile}, G., {Tenzer}, C., {Vacchi}, A., {Zampa}, G., {Zampa}, N., and
  {Zorzi}, N., ``{HERMES: An ultra-wide band X and gamma-ray transient monitor
  on board a nano-satellite constellation},'' {\em Nuclear Instruments and
  Methods in Physics Research A}~{\bf 936},  199--203 (Aug. 2019).

\bibitem{sanna20}
{Sanna}, A. et~al., ``{Timing techniques applied to distributed modular
  high-energy astronomy: the H.E.R.M.E.S. project},'' {\em Society of
  Photo-Optical Instrumentation Engineers (SPIE) Conference Series,}~{\bf
  11444-251} (2020).

\bibitem{fiore20}
{Fiore}, F. et~al., ``{The HERMES-Technologic and Scientific Pathfinder},''
  {\em Society of Photo-Optical Instrumentation Engineers (SPIE) Conference
  Series,}~{\bf 11444-166} (2020).

\bibitem{colagrossi20}
Colagrossi, A., Prinetto, J., Silvestrini, S., and Lavagna, M.~R., ``{Sky
  visibility analysis for astrophysical data return maximization in HERMES
  constellation},'' {\em Journal of Astronomical Telescopes, Instruments, and
  Systems}~{\bf 6}(4),  1 -- 25 (2020).

\bibitem{gatti84}
Gatti, E. and Rehak, P., ``Semiconductor drift chamber - an application of a
  novel charge transport scheme,'' {\em Nuclear Instruments and Methods in
  Physics Research}~{\bf 225}(3),  608 -- 614 (1984).

\bibitem{evangelista20}
{Evangelista}, Y. et~al., ``{The scientific payload on-board the HERMES-TP and
  HERMES-SP CubeSat missions,},'' {\em Society of Photo-Optical Instrumentation
  Engineers (SPIE) Conference Series,}~{\bf 11444-168} (2020).

\bibitem{marisaldi04}
{Marisaldi}, M., {Labanti}, C., and {Soltau}, H., ``{A pulse shape
  discrimination gamma-ray detector based on a silicon drift chamber coupled to
  a CsI(Tl) scintillator: prospects for a 1 keV-1 MeV monolithic detector},''
  {\em IEEE Transactions on Nuclear Science}~{\bf 51}(4),  1916--1922 (2004).

\bibitem{marisaldi08}
Marisaldi, M., Labanti, C., Fuschino, F., and Amati, L., ``{A broad energy
  range wide field monitor for next generation gamma-ray burst experiments},''
  {\em Nuclear Instruments and Methods in Physics Research Section A:
  Accelerators, Spectrometers, Detectors and Associated Equipment}~{\bf
  588}(1),  37 -- 40 (2008).
\newblock Proceedings of the First International Conference on Astroparticle
  Physics.

\bibitem{kamada12}
{Kamada}, K., {Yanagida}, T., {Pejchal}, J., {Nikl}, M., {Endo}, T.,
  {Tsutsumi}, K., {Fujimoto}, Y., {Fukabori}, A., and {Yoshikawa}, A.,
  ``{Crystal Growth and Scintillation Properties of Ce Doped ${\rm Gd}_{3}({\rm
  Ga},{\rm Al})_{5}{\rm O}_{12}$ Single Crystals},'' {\em IEEE Transactions on
  Nuclear Science}~{\bf 59}(5),  2112--2115 (2012).

\bibitem{dilillo20}
{Dilillo}, G. et~al., ``{A summary on an investigation of GAGG:Ce afterglow
  emission in the context of future space applications within the HERMES
  nanosatellite mission},'' {\em Society of Photo-Optical Instrumentation
  Engineers (SPIE) Conference Series,}~{\bf 11444-185} (2020).

\bibitem{ripa20}
{Ripa}, J. et~al., ``{A comparison of trapped particle models in Low Earth
  Orbit},'' {\em Society of Photo-Optical Instrumentation Engineers (SPIE)
  Conference Series,}~{\bf 11444-74} (2020).

\bibitem{campana20}
{Campana}, R. et~al., ``{The HERMES background and response simulations},''
  {\em Society of Photo-Optical Instrumentation Engineers (SPIE) Conference
  Series,}~{\bf 11444-248} (2020).

\bibitem{zampa11}
Zampa, G., Campana, R., Feroci, M., Vacchi, A., Bonvicini, V., {Del Monte}, E.,
  Evangelista, Y., Fuschino, F., Labanti, C., Marisaldi, M., Muleri, F.,
  Pacciani, L., Rapisarda, M., Rashevsky, A., Rubini, A., Soffitta, P., Zampa,
  N., Baldazzi, G., Costa, E., Donnarumma, I., Grassi, M., Lazzarotto, F.,
  Malcovati, P., Mastropietro, M., Morelli, E., and Picolli, L.,
  ``Room-temperature spectroscopic performance of a very-large area silicon
  drift detector,'' {\em Nuclear Instruments and Methods in Physics Research
  Section A: Accelerators, Spectrometers, Detectors and Associated
  Equipment}~{\bf 633}(1),  15 -- 21 (2011).

\bibitem{campana11}
Campana, R., Zampa, G., Feroci, M., Vacchi, A., Bonvicini, V., {Del Monte}, E.,
  Evangelista, Y., Fuschino, F., Labanti, C., Marisaldi, M., Muleri, F.,
  Pacciani, L., Rapisarda, M., Rashevsky, A., Rubini, A., Soffitta, P., Zampa,
  N., Baldazzi, G., Costa, E., Donnarumma, I., Grassi, M., Lazzarotto, F.,
  Malcovati, P., Mastropietro, M., Morelli, E., and Picolli, L., ``Imaging
  performance of a large-area silicon drift detector for x-ray astronomy,''
  {\em Nuclear Instruments and Methods in Physics Research Section A:
  Accelerators, Spectrometers, Detectors and Associated Equipment}~{\bf
  633}(1),  22 -- 30 (2011).

\bibitem{Rachevski14}
Rachevski, A., Zampa, G., Zampa, N., Campana, R., Evangelista, Y., Giacomini,
  G., Picciotto, A., Bellutti, P., Feroci, M., Labanti, C., Piemonte, C., and
  Vacchi, A., ``Large-area linear silicon drift detector design for x-ray
  experiments,'' {\em Journal of Instrumentation}~{\bf 9},  P07014--P07014 (jul
  2014).

\bibitem{campana16}
Campana, R., Fuschino, F., Labanti, C., Marisaldi, M., Amati, L., Fiorini, M.,
  Uslenghi, M., Baldazzi, G., Bellutti, P., Evangelista, Y., Elmi, I., Feroci,
  M., Ficorella, F., Frontera, F., Picciotto, A., Piemonte, C., Rachevski, A.,
  Rashevskaya, I., Rignanese, L.~P., Vacchi, A., Zampa, G., Zampa, N., and
  Zorzi, N., ``{A compact and modular x- and gamma-ray detector with a CsI
  scintillator and double-readout Silicon Drift Detectors},'' in [{\em Space
  Telescopes and Instrumentation 2016: Ultraviolet to Gamma
  Ray}{\nolinebreak\hspace{0.1em}]},  den Herder, J.-W.~A., Takahashi, T., and
  Bautz, M., eds.,  {\bf 9905},  1900 -- 1909, International Society for Optics
  and Photonics, SPIE (2016).

\bibitem{fuschino16}
Fuschino, F., Labanti, C., Campana, R., Gangemi, G.~C., Marisaldi, M.,
  Rignanese, L.~P., Baldazzi, G., Elmi, I., Evangelista, Y., Feroci, M., Zampa,
  G., Zampa, N., Rashevsky, A., Vacchi, A., Rashevskaya, I., Fabiani, S.,
  Zorzi, N., Piemonte, C., Bellutti, P., Ficorella, F., Borghi, G., and
  Burderi, L., ``{Characterization of a LaBr3 scintillator with multi-cell
  Silicon Drift Detector (SDD) readout},'' in [{\em Space Telescopes and
  Instrumentation 2016: Ultraviolet to Gamma Ray}{\nolinebreak\hspace{0.1em}]},
   den Herder, J.-W.~A., Takahashi, T., and Bautz, M., eds.,  {\bf 9905},  1910
  -- 1918, International Society for Optics and Photonics, SPIE (2016).

\bibitem{evangelista18}
Evangelista, Y., Ambrosino, F., Feroci, M., Bellutti, P., Bertuccio, G.,
  Borghi, G., Campana, R., Caselle, M., Cirrincione, D., Ficorella, F.,
  Fiorini, M., Fuschino, F., Gandola, M., Grassi, M., Labanti, C., Malcovati,
  P., Mele, F., Morbidini, A., Picciotto, A., Rachevski, A., Rashevskaya, I.,
  Sammartini, M., Zampa, G., Zampa, N., Zorzi, N., and Vacchi, A.,
  ``Characterization of a novel pixelated silicon drift detector ({PixDD}) for
  high-throughput x-ray astrophysics,'' {\em Journal of Instrumentation}~{\bf
  13},  P09011--P09011 (sep 2018).

\bibitem{cirrincione19}
Cirrincione, D., Ahangarianabhari, M., Ambrosino, F., Bajnati, I., Bellutti,
  P., Bertuccio, G., Borghi, G., Bufon, J., Cautero, G., Ceraudo, F.,
  Evangelista, Y., Fabiani, S., Feroci, M., Ficorella, F., Gandola, M., Mele,
  F., Orzan, G., Picciotto, A., Sammartini, M., Rachevski, A., Rashevskaya, I.,
  Schillani, S., Zampa, G., Zampa, N., Zorzi, N., and Vacchi, A., ``High
  precision mapping of single-pixel silicon drift detector for applications in
  astrophysics and advanced light source,'' {\em Nuclear Instruments and
  Methods in Physics Research Section A: Accelerators, Spectrometers, Detectors
  and Associated Equipment}~{\bf 936},  239 -- 241 (2019).
\newblock Frontier Detectors for Frontier Physics: 14th Pisa Meeting on
  Advanced Detectors.

\bibitem{bertuccio15}
Bertuccio, G., Ahangarianabhari, M., Graziani, C., Macera, D., Shi, Y.,
  Rachevski, A., Rashevskaya, I., Vacchi, A., Zampa, G., Zampa, N., Bellutti,
  P., Giacomini, G., Picciotto, A., and Piemonte, C., ``A silicon drift
  detector-{CMOS} front-end system for high resolution x-ray spectroscopy up to
  room temperature,'' {\em Journal of Instrumentation}~{\bf 10},
  P01002--P01002 (jan 2015).

\bibitem{Aha14}
Ahangarianabhari, M., Macera, D., Bertuccio, G., Malcovati, P., and Grassi, M.,
  ``Vega: A low-power front-end asic for large area multi-linear x-ray silicon
  drift detectors: Design and experimental characterization,'' {\em Nuclear
  Instruments and Methods in Physics Research Section A: Accelerators,
  Spectrometers, Detectors and Associated Equipment}~{\bf 770},  155 -- 163
  (2015).

\bibitem{campana14}
Campana, R., Evangelista, Y., Fuschino, F., Ahangarianabhari, M., Macera, D.,
  Bertuccio, G., Grassi, M., Labanti, C., Marisaldi, M., Malcovati, P.,
  Rachevski, A., Zampa, G., Zampa, N., Andreani, L., Baldazzi, G., Monte,
  E.~D., Favre, Y., Feroci, M., Muleri, F., Rashevskaya, I., Vacchi, A.,
  Ficorella, F., Giacomini, G., Picciotto, A., and Zuffa, M.,
  ``Characterization of the {VEGA} {ASIC} coupled to large area
  position-sensitive silicon drift detectors,'' {\em Journal of
  Instrumentation}~{\bf 9},  P08008--P08008 (aug 2014).

\bibitem{gandola2019lyra}
Gandola, M., Grassi, M., Mele, F., Malcovati, P., and Bertuccio, G., ``{LYRA: A
  Multi-Chip ASIC Designed for HERMES X and Gamma Ray Detector},'' in [{\em
  2019 IEEE Nuclear Science Symposium and Medical Imaging Conference
  (NSS/MIC)}{\nolinebreak\hspace{0.1em}]},   1--3, IEEE (2019).

\bibitem{grassi2020x}
Grassi, M., Gandola, M., Mele, F., Bertuccio, G., Malcovati, P., Fuschino, F.,
  Campana, R., Labanti, C., Fiorini, M., Evangelista, Y., et~al.,
  ``{X-/$\gamma$-Ray Detection Instrument for the HERMES Nano-Satellites Based
  on SDDs Read-Out by the LYRA Mixed-Signal ASIC Chipset},'' in [{\em 2020 IEEE
  International Instrumentation and Measurement Technology Conference
  (I2MTC)}{\nolinebreak\hspace{0.1em}]},   1--6, IEEE (2020).

\end{thebibliography}
\bibliographystyle{spiebib} 

\end{document}